\documentclass[reqno]{article}%12pt%
\usepackage{alltt}
\usepackage{graphicx}
\title{Self-interacting Electron as the Gauge Field Under the Ultimate Separation of the Absolute Quantum Motions}
\author{Peter Leifer}
\date{Or-Aqiva, Israel}
\begin{document}
\maketitle
\begin{abstract}
The problem of the reason of physical motion needs a review in the framework of quantum theory. The Aristotle's mistake, Galileo-Newton progress, Einstein physical geometry established the fundamental role of the spacetime geometry in the motion of fields and bodies. Quantum theory poses a new question about the motion of the quantum states and its reason in the quantum state space. The standard approach of quantum theory uses so-called method of the classical analogy where the action functional contains in the additive manner three terms: matter (free particles) + free fields + interaction term. Such approach leads to the quantum state space as some space of functions defined on the spacetime.
I think if one try to understand the peculiarity of the self-interacting quantum particles together with its ``field shell" then the classical scheme should be replaced. Then the role of the spacetime should be revised: the space of the unlocated pure quantum degrees of freedom and its geometry will play the fundamental role and the local dynamical spacetime arises as representation of the internal quantum motions (inverse representation).

I will discuss in this work a small but important change in the formulation of the field equations for the energy-momentum, orbital momentum and kinetic momentum of the self-interacting electron.

\end{abstract}


\vskip 0.1cm
\section{Introduction}
The old Poincar\'e idea on hypothetical stretches preventing the electron from the flying apart is alive. This way would be successful if the electron stability naturally connected with instability of the second and third lepton generations (muon and tauon).

I propose in this article a modified dynamical mechanism of the EM-like ``field shell" creation by the quantum electron. Attempts to use the affine gauge potential in the complex projective state space $CP(N-1)$ of the pure quantum degrees of freedom is known \cite{Le18/1,Le18/2,Le11,Le13,Le15,Le16} but the robust result was not achieved. Now I think the approach that I will discuss here will be prolific for future development. Namely, the role of the Jacobi field was clear for me as necessary element capable naturally involve the curvature of the $CP(N-1)$ in the new quantum field dynamics \cite{Le13} but there were difficulties with some technical details. The simple redefinition of the tangent vector field
$T^i = P^{\sigma}\Phi^i_{\sigma}+J^i_{\bot} = P^i+J^i_{\bot}$ instead of the old one $P^i = P^{\sigma}\Phi^i_{\sigma}$
serves as tangent vector to geodesic
$ T^i  = \frac{d \pi^i}{d S}= \lambda J^i_{\|} $
so that the covariant derivative of both sides vanishes \emph{identically}. Such definition intended to give rise to the compensation of the divergency of the geodesics in the vicinity of the ``north pole" $(\pi^1=\pi^2=\pi^3=0)$ of the $CP(3)$ and the electron stabilization by the gauge field $P^{\alpha}\Phi^i_{\alpha}$. One may think that the electron charged by the divergency of the transversal Jacobi field $J^i_{\bot}$ of the  geodesic variations in $CP(3)$ ``inflating" it due to the affine connection
\begin{eqnarray}
\Gamma^i_{kl} = \frac{1}{2}G^{ip^*} (\frac{\partial
G_{kp^*}}{\partial \pi^l} + \frac{\partial G_{p^*l}}{\partial
\pi^k}) = -  \frac{\delta^i_k \pi^{l^*} + \delta^i_l \pi^{k^*}}{1+
\sum |\pi^s|^2},
\end{eqnarray}\label{3}
and stabilized by the compensation field $P^{\sigma}\Phi^i_{\sigma}$ from the $AlgSU(4)$ so that their sum is proportional to the longitudinal Jacobi field and, hence, to the velocity $\frac{d \pi^i}{d S}= \lambda J^i_{\|}$  of the geodesic motion of the unlocated quantum state (UQS).

\section{The fundamental role of the state space}
Inertial motion does not requires some reason - it is simply exists in Nature at least in a good approximation, such is the postulate of the classical physics. More precisely, one may think that probably there are some reasons but they are unknown or not interesting for us. Einstein attempting to reject the preference of the inertial systems over accelerated frames explained why the classical formulation of the inertia principle is not satisfying \cite{Einstein_1921,Le16}. Quantum physics formally took into account the Poincar\'e symmetry but it faces with essential difficulties \cite{Weinberg_2017} that rooted in basic foundations of two theories \cite{Oxford_Q}.
Now we should return to the problem of the inertial motion of a single ``elementary" quantum particle like electron.

I have wrote that acceleration is only an ``external" exhibit of the non-inertial motion:
the \emph{deformation} of a body or internal quantum states is most deep result of the interaction \cite{Le16}. The quantum version of the classical gauge theory of the finite deformations of the ``unlocated shape" of a body is interesting for us \cite{Littlejohn,Le13,Le15,Le16,Le18/1,Le18/2}.

Attempt to find the ``peaceful coexistence" of relativity and quantum laws in the manner of the intrinsic unification, i.e. starting from the pure quantum degrees of freedom was called ``Quantum Relativity" \cite{Le18/1,Le18/2}.
Such intrinsic unification of the quantum theory and relativity is possible only on the way of the serious deviation from traditional assumptions about a priori spacetime structure and the Yang-Mills generalization of the well known $U(1)$ Abelian gauge symmetry of the classical electrodynamics. More general gauge theory should be constructed as the quantum version of the gauge theory of the deformable bodies - the gauge theory of the deformable unlocated quantum states (UQS's). This means that localization of quantum state is achievable in a functional space since the distance between quantum states is strictly defined value whereas the distance between bodies (particle) is an approximate value, at best \cite{Le13,Le15,Le16,Le18/1,Le18/2}. Thereby, all well known solid frames and clocks even with the corrections of special relativity should be replaced by the flexible and anholonomic quantum setup.
The Yang-Mills arguments about the spacetime coordinate dependence of the gauge unitary rotations should be reversed on the  dependence of the spacetime structure on the unholonomy of the gauge transformations of the flexible quantum setup.

The appearance of the geometric gauge fields is the well known phenomenon in the wide area of the ``geometric phase" \cite{Aitchison,Littlejohn}. These fields frequently connected with some a singularity of the mapping. But the fundamental physical fields cannot have a singular source. I try connect EM-like field of the quantum electron with the curvature of the coset sub-manifold of the unitary group acting on the space of the unlocated quantum states.

There is an obvious fact: trajectory of classical particle in spacetime is merely an idealization as well as ``free field configuration". Therefore the independent variation of such classical elements as trajectories or potentials is an approximate too.
Only internal pure quantum degrees of freedom common for the quantum ``gauge fields" and the ``fields of matter"  subject to the independent variations. Variation of the UQS's should lead to the ``field particle", i.e. to the quantum particle together with its ``field shell". There is no the classical separation of the ``particle motion" and the ``field equations" under the independent variations of the particle trajectories and potentials.

\section{Quantum Relativity}
The principle of Quantum Relativity (QR) assumes the invariance of physical properties of ``quantum particles" i.e. their quantum numbers like mass, spin, charge, etc. in any conceivable quantum setup or ambient. Such invariance may be lurked, say, behind two amplitudes $|\Psi_1>, |\Psi_2>$ in two different quantum setups $S_1$ and $S_2$.  The invariant content of these properties will be discussed here under the infinitesimal variation of the ``flexible quantum setup" described by the amplitudes $|\Psi(\pi,P)>$ due to a small variation of the boson electromagnetic-like field $P^{\sigma}(\pi)$ treated as \emph{the set of the scalar functions} relative ${\pi^i}$ coordinates in $CP(N-1)$. The DST dependence of $P^{\sigma}(\pi)$ will be established after the separation of the shifts, boosts and rotations in the manifold of the $SU(N)$ generators.

The mathematical formulation of the QR principle is based on the \emph{similarity} of any physical systems which are built on the ``elementary" particles. This similarity is obvious only on the level of the pure quantum degrees of freedom of quantum particles. Therefore, all ``external" details of the ``setup" should be discarded as non-essential and only the relations of components of the ``unitary spin" like $(\pi^1=\frac{\psi^2}{\psi^1},...,\pi^{N-1}=\frac{\psi^{N}}{\psi^1})$  should be taken into account. \emph{These relations will be assumed Lie-dragged during global unitary transformations in $C^N$ and they are taken as the local projective coordinates in the complex projective Hilbert space $CP(N-1)$.} One may think about these coordinates as parameters of the ``shape of quantum particle" in the spirit of the \cite{Littlejohn}.
On the other hand, the local projective coordinates $\pi^i$ in $CP(N-1)$ taking the place of the ``basic particles" like ``goldstone bosons" in the method of the phenomenological Lagrangians \cite{Volkov} and the Jacobi fields serve as the source of the EM-like fields in the DST due to the affine gauge potential in $CP(N-1)$.

\section{The coset state space, deformation of quantum state}
The fundamental quantum degrees of freedom like spin, charge, hyper-charges, etc., are common for gauge and matter fields. These fundamental quantum motions take the place in the manifold of the UQS's which described by the rays of states $|\psi> \in C^N$ of the ``unitary spin" $S: 2S+1=N$ . Physics requires to use in this background the local coordinates of UQS's and the state-dependent generators of the unitary group $G=SU(N)$ \cite{Le04}. This nonlinear representation of the $SU(N)$ group on the coset manifold $G/H=SU(N)/S[U(1) \times U(N-1)]=CP(N-1)$ is primary and this is independent on the spacetime manifold. The last one should be introduced in a special section of the fiber bundle over $CP(N-1)$ \cite{Le13,Le15,Le16,Le18/1,Le18/2}. The breakdown of the global $SU(N)$ symmetry down to the isotropy subgroup $H_{|\psi>}=U(1) \times U(N-1)$ of the some quantum state $|\psi>$ has natural geometric counterpart in $CP(N-1)$.

The coset manifold $G/H_{|\psi>}=SU(N)/S[U(1) \times U(N-1)]=CP(N-1)$ contains locally unitary transformations \emph{deforming} ``initial" quantum state $|\psi>$. This means that $CP(N-1)$ contains physically distinguishable, ``deformed" quantum states. Thereby the unitary transformations from $G=SU(N)$ of the basis in the Hilbert space may be identified with the unitary state-dependent gauge field $U(|\psi>)$ that may be represented by the $N^2-1$ unitary generators as functions of the local projective coordinates $(\pi^1,...,\pi^{N-1})$ \cite{Le13}. This manifold resembles the ``shape space" of the deformable body \cite{Littlejohn,Le13,Le15,Le16,Le18/1,Le18/2}. But now it is the manifold of the deformed physically distinguishable UQS's, i.e. the geometric, invariant counterpart of the quantum interaction or self-interaction. Then the classical acceleration is merely an ``external" consequence of this complicated quantum dynamics in the some section of the frame fiber bundle over $CP(N-1)$. The local dynamical variables (LDV's) are new essential elements of the new quantum dynamics \cite{Le04}. They should be expressed in terms of the local coordinates $\pi^k$ of UQS's. Thereby they
will live in the geometry of $CP(N-1)$ with the Fubini-Study metric tensor
\begin{equation}
G_{ik^*} = (1/{\kappa})[(1+ \sum |\pi^s|^2) \delta_{ik}- \pi^{i^*} \pi^k](1+
\sum |\pi^s|^2)^{-2},
\end{equation}\label{2}
where $\kappa$ is holomorphic sectional curvature of the $CP(N-1)$ \cite{KN}.
The contra-variant metric tensor field
\begin{equation}
G^{ik^*} =\kappa (\delta^{ik} + \pi^{i} \pi^{k*})(1+
\sum |\pi^s|^2),
\end{equation}\label{3}
is inverse to the $G_{ik^*}$ thereby
\begin{equation}
G_{ik^*}G^{i^*q} = \delta_{k}^q.
\end{equation}\label{3}

The flexible quantum setup inherently connected with local projective coordinates will be built from so-called LDV's \cite{Le04}. These LDV's realize
a non-linear representation of the unitary global $SU(N)$ group in
the Hilbert state space $C^N$. Namely, $N^2-1$ generators of $G =
SU(N)$ may be divided in accordance with the Cartan decomposition:
$[B,B] \in H, [B,H] \in B, [B,B] \in H$. The $(N-1)^2$ generators
\begin{eqnarray}
\Phi_h^i \frac{\partial}{\partial \pi^i}+c.c. \in H,\quad 1 \le h
\le (N-1)^2
\end{eqnarray}\label{4}
of the isotropy group $H = U(1)\times U(N-1)$ of the ray (Cartan
sub-algebra) and $2(N-1)$ generators
\begin{eqnarray}
\Phi_b^i \frac{\partial}{\partial \pi^i} + c.c. \in B, \quad 1 \le b
\le 2(N-1)
\end{eqnarray}\label{5}
are the coset $G/H = SU(N)/S[U(1) \times U(N-1)]$ generators
realizing the breakdown of the $G = SU(N)$ symmetry.
Notice, the partial derivatives are defined here as usual: $\frac{\partial }{\partial \pi^i} = \frac{1}{2}
(\frac{\partial }{\partial \Re{\pi^i}} - i \frac{\partial }{\partial
\Im{\pi^i}})$ and $\frac{\partial }{\partial \pi^{*i}} = \frac{1}{2}
(\frac{\partial }{\partial \Re{\pi^i}} + i \frac{\partial }{\partial
\Im{\pi^i}})$.

Here $\Phi^i_{\sigma}, \quad 1 \le \sigma \le N^2-1 $
are the coefficient functions of the generators of the non-linear
$SU(N)$ realization. They give the infinitesimal shift of the
$i$-component of the generalized coherent state driven by the $\sigma$-component
of the unitary  field $\exp(i\epsilon \lambda_{\sigma})$ rotating by the
generators of $Alg SU(N)$ and they are defined as follows:
\begin{equation}
\Phi_{\sigma}^i = \lim_{\epsilon \to 0} \epsilon^{-1}
\biggl\{\frac{[\exp(i\epsilon \lambda_{\sigma})]_m^i \psi^m}{[\exp(i
\epsilon \lambda_{\sigma})]_m^j \psi^m }-\frac{\psi^i}{\psi^j} \biggr\}=
\lim_{\epsilon \to 0} \epsilon^{-1} \{ \pi^i(\epsilon
\lambda_{\sigma}) -\pi^i \},
\end{equation}\label{6}
\cite{Le13}.

\section{Does the dynamical instability of the Jacobi field generate mass, electric charge and spin?}
My fundamental assumption is that the physically essential deformation of the internal quantum state is the process of motion of UQS along the geodesic in $CP(N-1)$. Then the very narrow class of deformations is the class of the geodesic-to-geodesic variations associated with the Jacobi fields. One may look on the dynamical problem from the point of view of the ``the control optimization" where unitary field $SU(N)$ of the chiral type rotates the local frame in $AlgSU(N)$ so that UQS moves along geodesic in $CP(N-1)$ with variable energy-momentum in the flexible 10D DST (a la S. Dali ``The Persistence of Memory").

I propose the dynamical model of the electric charge as the consequence of the dynamics of UQS of the electron. The Jacobi equation for the geodesic in $CP(N-1)$ looks as follows:
\begin{eqnarray}
\frac{d^2 J^i}{dS^2}+2\Gamma^i_{kl}\frac{d J^k}{dS}\frac{d \pi^l}{dS}+R^i_{klm^*}J^k
\frac{d \pi^l}{dS}\frac{d \pi^{m^*}}{dS}=0,
\end{eqnarray}\label{}
where $dS^2 = G_{ik^*}d\pi^i d\pi^{k^*}$.
This equation being written in the reference frame parallel transported along geodesic has very simple solutions \cite{Besse}. The parallel transported functional reference frame is an analog of the ``freely falling down" system. But such orthogonal reference frame sharply differs from the reference frame dictated by physics. For self-interacting quantum electron the reference frame in $CP(3)$ was built from the LDV's vector fields associated with the four Dirac matrices from $AlgSU(4)$ responsible for shifts in the DST and six matrices responsible for the boosts and rotations in 10D DST. Totally eleven $\lambda$- matrices of $AlgSU(4)$ from the fifteen 15 have been used. Last investigation shows that our field equations for momentum, angular momentum, and kinematic momentum should be reformulated in following manner. The requirement is as before: the tangent vector to the curve in $CP(3)$ should be parallel transported, i.e. the the curve should be geodesic in the $CP(3)$ but the velocity of the traversing should be variable. In order to obey this condition I define new vector field
\begin{eqnarray}
T^i = P^{\sigma}\Phi^i_{\sigma}+J^i_{\bot}= \lambda J^i_{\|}.
\end{eqnarray}\label{}
so that $P^{\sigma}\Phi^i_{\sigma}$ compensate the instability of $J^i_{\bot}$ as seen from Jacobi equation (8) \cite{Besse}. Then one has the identically
\begin{eqnarray}
T^i_{;k} = [P^{\sigma}\Phi^i_{\sigma}+J^i_{\bot}]_{;k}= \lambda [J^i_{\|}]_{;k}=0.
\end{eqnarray}\label{}
This procedure introduces a modified affine gauge field with variations in the DST
instead of the transition to the parallel transported frame. The Jacobi equation in such frame has the narrow class of geodesic traversing with a constant velocity. The original equation (8) describes more general Jacobi fields with longitudinal and transversal variations of the traversing speed.

Let me take the typical geodesic line in $CP(N-1)$ in terms of the local coordinates $\pi^i$;
\begin{equation}\label{}
\pi^i=\frac{f^i}{g}\tan{g\tau}
\end{equation}
where $g=\sqrt{\sum_{s=1}^{s=N-1}|f^s|^2}$. Along such the geodesic one has following expressions for affine connection
\begin{eqnarray}
\Gamma^i_{kl} = \frac{1}{2}G^{ip^*} (\frac{\partial
G_{kp^*}}{\partial \pi^l} + \frac{\partial G_{p^*l}}{\partial
\pi^k}) = -  \frac{\delta^i_k \pi^{l^*} + \delta^i_l \pi^{k^*}}{1+
\sum |\pi^s|^2} \cr
= -\frac{\delta^i_lf^{k*}+\delta^i_kf^{l*}}{g}\sin{g\tau}\cos{g\tau},
\end{eqnarray}\label{3}
and the curvature tensor
\begin{eqnarray}
R^i_{klm^*} = \kappa^2(\delta^i_l G_{km^*} + \delta^i_k G_{lm^*})\cr
= \kappa( \delta^i_l   \delta_{km} +    \delta^i_k   \delta_{lm}
 -\frac{\delta^i_lf^{k*}f^m + \delta^i_k f^{l*}f^m}{g^2}\sin{g\tau}^2 )\cos{g\tau}^2.
\end{eqnarray}\label{3}
Therefore, second and third terms in the Jacobi equation (8)
is as follows:
\begin{eqnarray}
\Gamma^i_{kl} \frac{d \pi^l}{d\tau}=
-g \tan{g\tau}(\delta^i_k + \frac{f^i f^{k*}}{g^2}),
\end{eqnarray}\label{3}
and
\begin{eqnarray}
R^i_{klm^*}\frac{d \pi^l}{d\tau}\frac{d \pi^{m*}}{d\tau}=\kappa g^2 (\delta^i_k + \frac{f^i f^{k*}}{g^2}).
\end{eqnarray}
One has the second order linear homogeneous differential equation
\begin{eqnarray}
\frac{d^2 J^i}{d\tau^2}-2g\tan{g\tau}(\delta^i_k + \frac{f^i f^{k*}}{g^2})\frac{d J^k}{d\tau} +\kappa g^2 (\delta^i_k + \frac{f^i f^{k*}}{g^2})J^k
=0.
\end{eqnarray}\label{}
Taking for simplicity the set $(f^1=1,f^2=f^3=0)$ one get the three equations
\begin{eqnarray}
\frac{d^2 J^1}{d\tau^2}-4\tan{g\tau}\frac{d J^1}{d\tau} +2\kappa J^1=0,\cr
\frac{d^2 J^2}{d\tau^2}-2\tan{g\tau}\frac{d J^2}{d\tau} +\kappa J^2=0,\cr
\frac{d^2 J^3}{d\tau^2}-2\tan{g\tau}\frac{d J^3}{d\tau} +\kappa J^3=0.
\end{eqnarray}\label{}
The general solutions of these equations are as follow:
\begin{eqnarray}
J^1 = C_1\cos(\tau)^{-3/2}P(\sqrt{2\kappa+4}-1/2, 3/2, \sin(\tau)) \cr
+ C_2 \cos(\tau)^{-3/2}Q(\sqrt{2\kappa+4}-1/2, 3/2, \sin(\tau)),\cr
J^2 = C_3\cos(\tau)^{-1}\sinh(\sqrt{-1-\kappa}\tau))
+ C_4 \cos(\tau)^{-1}\cosh(\sqrt{-1-\kappa}\tau)),\cr
J^3 = C_5\cos(\tau)^{-1}\sinh(\sqrt{-1-\kappa}\tau))
+ C_6 \cos(\tau)^{-1}\cosh(\sqrt{-1-\kappa}\tau)),
\end{eqnarray}\label{}
where $P(\sqrt{2\kappa+4}-1/2, 3/2, \sin(\tau)),Q(\sqrt{2\kappa+4}-1/2, 3/2, \sin(\tau))$ are the associate Legendre functions of the first and the second kinds.
It is clear that more complicated choice for the complex velocity traversing of the basic geodesic $(f^1,f^2,f^3,...,f^{N-1})$ gives more complicated solutions. Such solutions should be included in the equations (10).

\section{How the field motions in DST created by the UQS motions in $CP(3)$?}
The old problem of the accelerated charged particle is an acute challenge for QFT, high energy physics, and for the theory of elementary particles. The problem of the self-interaction and, hence, self-acceleration must be formulated now in terms of the internal QDF's.

I assumed that the reason of the inertial motion of the self-interacting electron may be described by the internal motions of the QDF's in $CP(3)$. This means that dynamical shifts, rotations and boosts may be represented by the Poincar\'e generators expressed as the special linear combinations of the Lie derivatives of the local projective coordinates $(\pi^1,\pi^2,\pi^3)$ in the directions given by the Dirac matrices in the Weyl representation and the six additional matrices of $AlgSU(4)$. This construction is most transparent for the fundamental fermion like the electron. More general case of higher dimension should be discussed elsewhere. Probably, after all it is possible to identify the quantum electron itself together with its ``field shell" with dynamical shifts, rotations and boosts  in the intriguer manner: vector fields of shifts are identical to the components of the energy-momentum plus four-potential, boosts identical to the components of electric-like field, and rotations identical to the components of the magnetic-like field. But this possibility requires additional investigation.

I will use the following set of the Dirac matrices
\begin{eqnarray}\label{43}
\gamma_t=\left( \begin {array}{cccc}
0&0&1&0 \cr
0&0&0&1 \cr
1&0&0&0 \cr
0&1&0&0
\end {array} \right),
\gamma_1=-i \sigma_1=\left( \begin {array}{cccc}
0&0&0&-1 \cr
0&0&-1&0 \cr
0&1&0&0 \cr
1&0&0&0
\end {array} \right), \cr
\gamma_2=-i \sigma_2=\left( \begin {array}{cccc}
0&0&0&i \cr
0&0&-i&0 \cr
0&-i&0&0 \cr
i&0&0&0
\end {array} \right),
\gamma_3=-i \sigma_3=\left( \begin {array}{cccc}
0&0&-1&0 \cr
0&0&0&1 \cr
1&0&0&0 \cr
0&-1&0&0
\end {array} \right).
\end{eqnarray}
Then the corresponding coefficients of the $SU(4)$ generators will be calculated according to the equation
\begin{equation}
\Phi_{\mu}^i = \lim_{\epsilon \to 0} \epsilon^{-1}
\biggl\{\frac{[\exp(i\epsilon \gamma_{\mu})]_m^i \psi^m}{[\exp(i
\epsilon \gamma_{\mu})]_m^j \psi^m }-\frac{\psi^i}{\psi^j} \biggr\}=
\lim_{\epsilon \to 0} \epsilon^{-1} \{ \pi^i(\epsilon
\gamma_{\mu}) -\pi^i \},
\end{equation}\label{6}
\cite{Le13}
that gives
\begin{eqnarray}
\Phi_{0}^1(\gamma_{t})&=&i(\pi^3-\pi^1 \pi^2), \quad \Phi_{0}^2(\gamma_{t})=i(1-(\pi^2)^2),
\quad \Phi_{0}^3(\gamma_{t})=i(\pi^1-\pi^2 \pi^3); \cr
\Phi_{1}^1(\gamma_{1})&=&-i(\pi^2 -\pi^1 \pi^3),
\quad \Phi_{1}^2(\gamma_{1})=-i(-\pi^1 -\pi^2 \pi^3),
\quad \Phi_{1}^3(\gamma_{1})=-i(-1 -(\pi^3)^2); \cr
\Phi_{2}^1(\gamma_{2})&=&-i(i(\pi^2 +\pi^1 \pi^3)),
\quad \Phi_{2}^2(\gamma_{2})=-i(i(\pi^1 +\pi^2 \pi^3)),
\quad \Phi_{2}^3(\gamma_{2})=-i(i(-1 +(\pi^3)^2)); \cr
\Phi_{3}^1(\gamma_{3})&=&-i(-\pi^3 -\pi^1 \pi^2),
\quad \Phi_{3}^2(\gamma_{3})=-i(-1 -(\pi^2)^2),
\Phi_{3}^3(\gamma_{3})=-i(\pi^1 -\pi^2 \pi^3).
\end{eqnarray}\label{15}
Such choice of the vector fields leads to the ``imaginary" basic in local DST which conserves $4D$ Eucledian geometry along geodesic in $CP(3)$ for real four vectors $(p^0,p^1,p^2,p^3)$ and correspondingly $4D$ pseudo-Eucledian geometry for four vectors $(ip^0,p^1,p^2,p^3)$.

The complex DST ``tangent vector" in $\mu$ direction defines the four complex shifts in DST that will be introduced as follows:
\begin{eqnarray}
\frac{\partial }{\partial x^{\mu}} = \Phi_{\mu}^i \frac{\partial }{\partial \pi^i}
\end{eqnarray}\label{}
for $0\leq \mu \leq 3$. In fact one may define the similar ``tangent vector" in $\sigma$ direction
\begin{eqnarray}
\frac{\partial }{\partial x^{\sigma}} = \Phi_{\sigma}^i \frac{\partial }{\partial \pi^i}
\end{eqnarray}\label{}
for $1 \leq \sigma \leq 15$ in the space $R^{15}$ of the adjoint representation of the $SU(4)$. Thereby, the DST cannot be treated as the ``space of events". It is rather 10-dimension subspace of the adjoint representation of the $SU(4)$. The quantum operator of the energy-momentum will be expressed as the shift operator
\begin{eqnarray}
\vec{P}_{\mu}=i\hbar \frac{\partial }{\partial x^{\mu}} = i\frac{\hbar}{L}  \Phi_{\mu}^i \frac{\partial }{\partial \pi^i}.
\end{eqnarray}\label{}
Now one may introduce six generators of the boosts and rotations started from the well known definitions in terms of Dirac matrices \cite{Feynman} where I put
$1 \leq \alpha \leq 3$.

\begin{eqnarray}\label{43}
B_x=(i/2)\gamma_t \gamma_x=(i/2) \left( \begin {array}{cccc}
0&1&0&0 \cr
1&0&0&0 \cr
0&0&0&-1 \cr
0&0&-1&0
\end {array} \right),\cr
B_y=(i/2) \gamma_t \gamma_y=(i/2)\left( \begin {array}{cccc}
0&i&0&0 \cr
-i&0&0&0 \cr
0&0&0&-i \cr
0&0&i&0
\end {array} \right), \cr
B_z=(i/2) \gamma_t \gamma_z=(i/2)\left( \begin {array}{cccc}
1&0&0&0 \cr
0&-1&0&0 \cr
0&0&-1&0 \cr
0&0&0&1
\end {array} \right),\cr
R_x=(i/2) \gamma_y \gamma_z= (i/2) \left( \begin {array}{cccc}
0&i&0&0 \cr
i&0&0&0 \cr
0&0&0&i \cr
0&0&i&0
\end {array} \right),\cr
R_y=(i/2) \gamma_z \gamma_x= (i/2) \left( \begin {array}{cccc}
0&-1&0&0 \cr
1&0&0&0 \cr
0&0&0&-1 \cr
0&0&1&0
\end {array} \right),\cr
R_z=(i/2) \gamma_x \gamma_y= (i/2) \left( \begin {array}{cccc}
i&0&0&0 \cr
0&-i&0&0 \cr
0&0&i&0 \cr
0&0&0&-i
\end {array} \right).
\end{eqnarray}
Using the modified definition (20) one may find the corresponding coefficient functions of the vector fields of the Lorentz generators for boosts
\begin{equation}
\Phi^i(B_\alpha) = \lim_{\epsilon \to 0} \epsilon^{-1}
\biggl\{\frac{[\exp(\epsilon B_{\alpha})]_m^i \psi^m}{[\exp(
\epsilon B_{\alpha})]_m^j \psi^m }-\frac{\psi^i}{\psi^j} \biggr\}=
\lim_{\epsilon \to 0} \epsilon^{-1} \{ \pi^i(\epsilon
B_{\alpha}) -\pi^i \},
\end{equation}\label{6}
\begin{eqnarray}
\Phi^1(B_x)=\frac{1}{2}(1-(\pi^1)^2),
\Phi^2(B_x)=\frac{-1}{2}(\pi^3+\pi^1\pi^2),
\Phi^3(B_x)=\frac{-1}{2}(\pi^2+\pi^1\pi^3),\cr
\Phi^1(B_y)=-\frac{i}{2}(1+(\pi^1)^2),
\Phi^2(B_y)=-\frac{i}{2}(\pi^3+\pi^1\pi^2),
\Phi^3(B_y)=\frac{i}{2}(\pi^2-\pi^1\pi^3),\cr
\Phi^1(B_z)=-\pi^1,
\Phi^2(B_z)=-\pi^2,
\Phi^3(B_z)=0,
\end{eqnarray}
and rotations
\begin{equation}
\Phi^i(R_\alpha) = \lim_{\epsilon \to 0} \epsilon^{-1}
\biggl\{\frac{[\exp(\epsilon R_{\alpha})]_m^i \psi^m}{[\exp(
\epsilon R_{\alpha})]_m^j \psi^m }-\frac{\psi^i}{\psi^j} \biggr\}=
\lim_{\epsilon \to 0} \epsilon^{-1} \{ \pi^i(\epsilon
R_{\alpha}) -\pi^i \},
\end{equation}\label{6}

\begin{eqnarray}
\Phi^1(R_x)=\frac{i}{2}(1-(\pi^1)^2),
\Phi^2(R_x)=\frac{i}{2}(\pi^3-\pi^1\pi^2),
\Phi^3(R_x)=\frac{i}{2}(\pi^2-\pi^1\pi^3),\cr
\Phi^1(R_y)=\frac{1}{2}(1+(\pi^1)^2),
\Phi^2(R_y)=-\frac{1}{2}(\pi^3-\pi^1\pi^2),
\Phi^3(R_y)=\frac{1}{2}(\pi^2+\pi^1\pi^3),\cr
\Phi^1(R_z)=-i\pi^1,
\Phi^2(R_z)=0,
\Phi^3(R_z)=-i\pi^3,
\end{eqnarray}
Thereby, the eight $\lambda$-matrices  $(\lambda_4,\lambda_{11}),(\lambda_2,\lambda_{14}),(\lambda_1,\lambda_{13}),(\lambda_5,\lambda_{12})$ of the $AlgSU(4)$ were involved in the definition of the shift vector fields associated with the inertial terms and the four-potentials. It is easy to see that additional diagonal matrices $,(\lambda_{3}),(\lambda_{8}),(\lambda_{15})$ must be involved into the boosts
\begin{eqnarray}
\vec{B}_{\alpha}=  \Phi^i(B_{\alpha}) \frac{\partial }{\partial \pi^i}
\end{eqnarray}\label{}
and rotations
\begin{eqnarray}
\vec{R}_{\alpha}=  \Phi^i(R_{\alpha}) \frac{\partial }{\partial \pi^i}.
\end{eqnarray}\label{}
generators. The commutators of these vector fields may be found in \cite{Le18/2}.

\section{New field equations}
In order to find physically acceptable solutions of the equation (10) one needs to put the gauge and the ``boundary" restrictions on meanwhile undefined functions $P^{\sigma}$.
It is worse while to recall that EM potential by itself serves as an analog of the ``border" what was initially strange for E. Schr\"odinger \cite{Schr_28}. However, Schr\"odinger had known Coulomb potential whereas we need to find more general solution with inertial term and modified non-singular EM-like potentials.

Our requirement tells that the projection of the trajectory of a single quantum particle onto $CP(N-1)$ should be a geodesic. Hence, the covariant derivative in the sense of the Fubini-Study metric of the velocity of UQS $\frac{d \pi^i}{d \tau}$ should be zero
\begin{eqnarray}\label{43}
( P^{\sigma}\Phi_{\sigma}^i)_{;k} + J^i_{\bot;k}= \frac{\partial P^{\sigma}}{\partial \pi^k}\Phi_{\sigma}^i + P^{\sigma} (\frac{\partial \Phi_{\sigma}^i}{\partial \pi^k}+
\Gamma^i_{kl} \Phi_{\sigma}^l) + J^i_{\bot;k} = 0.
\end{eqnarray}
One sees that the dynamical system for non-linear field momentum is self-consistent since the speed of the traversing the geodesic in $CP(N-1)$ is not a constant but a variable value ``modulated" by the field coefficients $P^{\sigma}$.

In general case of the full Poincar\'e motions in $10D$ DST one has the equation
\begin{eqnarray}\label{43}
 \frac{\partial P^{\mu}}{\partial x^{\mu}} + P^{\mu} (\frac{\partial \Phi_{\mu}^i}{\partial \pi^i} +
\Gamma^i_{il} \Phi_{\mu}^l) +
\frac{\partial K^{\alpha}}{\partial u^{\alpha}} + K^{\alpha} (\frac{\partial \Phi^i(B_{\alpha})}{\partial \pi^i} +
\Gamma^i_{il} \Phi^l(B_{\alpha})) \cr +
\frac{\partial M^{\alpha}}{\partial {\omega}^{\alpha}} + M^{\alpha} (\frac{\partial \Phi^i(R_{\alpha})}{\partial \pi^i} +
\Gamma^i_{il} \Phi^l(R_{\alpha})) + J^i_{\bot;i} = 0.
\end{eqnarray}
with wide class of the TWS's. The DST argument of the TWS function $\xi=\frac{1}{\hbar}q_{a}C^{a}, (1\leq a \leq 10)$ will be equal in some approximation to the action invariant of the single classical material point
\begin{eqnarray}
 S = -a_{\mu}P^{\mu} + \frac{1}{2}\Omega_{\mu \nu} M^{\mu \nu} =const
\end{eqnarray}
under the appropriate choice of these constants. The choice of the physically acceptable solution depends on the formulation of the ``boundary problem" in the functional space over $CP(3)$. It is not formulated yet properly. One may assume that the ``Schr\"odinger equation" with the relativistic Hamiltonian vector field
\begin{eqnarray}
\vec{H}=  c[P^{\mu} \Phi_{\mu}^i +K^{\alpha}\Phi^i(B_{\alpha})
 +M^{\alpha}\Phi^i(R_{\alpha})+ J^i_{\bot}] \frac{\partial }{\partial \pi^i} + c.c.
\end{eqnarray}\label{}
may be used for the eigen-value problem in terms of the PDE for the total wave function.
Then the speed of the UQS components should be satisfied the following
equation of characteristics
\begin{eqnarray}
 \frac{d \pi^i}{d \tau} =  \frac{c}{\hbar}[P^{\mu} \Phi_{\mu}^i+K^{\alpha}\Phi^i(B_{\alpha})
 +M^{\alpha}\Phi^i(R_{\alpha})+ J^i_{\bot}]
\end{eqnarray}\label{}
where $\tau$ is the \emph{quantum elapsed time counted from the start of the internal motion}. Such quantum internal motion is absolute even for free quantum electron.
The Hamiltonian vector field leads to the quasi-linear PDE ``Schr\"odinger equation"
\begin{eqnarray}\label{43}
i\hbar \frac{d \Psi(\pi,q,p)}{d\tau} = [cP^{\mu} \Phi_{\mu}^i+K^{\alpha}\Phi^i(B_{\alpha})
 +M^{\alpha}\Phi^i(R_{\alpha})+ J^i_{\bot}]\frac{\partial \Psi(\pi,q,p)}{\partial \pi^i} \cr + c.c.= E[\Psi(\pi,q,p)] \Psi(\pi,q,p),
\end{eqnarray}
where the coordinates $(p,q)$ correspond to the shifts, rotations, boosts and gauge parameters of the local DST, and $E[\Psi(\pi,q,p)]$ is a functional of the total quantum state. Since the all geodesics in $CP(N-1)$ are closed and $\pi$-periodic the natural quantization may be applied to the each from the four components of the total wave function in the areas
$(U_1:\{\psi_1 \neq 0\},U_2:\{\psi_2 \neq 0\},U_3:\{\psi_3 \neq 0\},U_4:\{\psi_4 \neq 0\}) $.

\section{Conclusion}
Intrinsic unification of the relativity and quantum physics requires ultimately separate the absolute motion of the unlocated quantum states of the pure quantum degrees of freedom in the quantum particles like electron from the ``external", even inertial motion.

P. Dirac proposed the dynamical model of the spin-less electron where the classical Coulomb repulsive force was compensated by the surface tension \cite{Dirac_62}.
In such a model the muon looks as radial oscillations of the electron. The predicted mass of the muon was about $53m_e$. The modern topological model of the electric charges was promising but it has obvious difficulties with the prediction of the mass relations in the lepton generation \cite{Faber}.

In the framework of the Quantum Relativity I discussed a new kind of the gauge theory
of the extended quantum electron. In such theory the origin of the electric repulsive force is rooted in the affine gauge potential that makes the ``north pole" of the unlocated quantum state space $CP(3)$ unstable for the ordinary Jacobi vector field. The compensation field stabilizing the electron from the flying apart was found in the anholonomic frame of the vector fields in the $AlgSU(4)$. The picture looks as follows: the divergency of the Jacobi field (proportional to the electric charge) pushes UQS down to the valley of the affine gauge potential along basic geodesic and three complex coset components of the compensation field decelerate this motion, the same time nine real components of the isotropy subgroup $H=U(1)\times U(3)$ should lead to the spin of the electron. Definitely, it is only hope! All equations have the analytic solutions but the hard problem of the ``boundary conditions" in the functional space is the obstacle for the quantity estimations. But it is clear that the calculation of the ``self-acceleration" and acceleration $\frac{\partial P^{\alpha}}{m \partial x^0}$ may essentially differs from the well known problematic divergency and the ``runaway solution".

\end{document}